\documentstyle[12pt]{article}
\begin{document}
\def\square{\hbox{\vrule\vbox{\hrule\phantom{o}\hrule}\vrule}}\renewcommand{\thefootnote}{\fnsymbol{footnote}}
\def\wh{\widehat}
\begin{center}
{\bf SELF-DUAL SPIN-3 AND 4 THEORIES} \\ [7mm]
C. Aragone \\
Departamento de F\'{\i}sica, Universidad Sim\'on Bol\'{\i}var, \\
Apartado 8900, Caracas 1080A, Venezuela \\ [4mm]
and \\ [4mm]
A. Khoudeir \footnote{Permanent address: Departamento de F\'{\i}sica,
Facultad de Ciencias, Universidad de los Andes, Apartado 5100, M\'erida,
Venezuela}\\
International Centre For Theoretical Physics, Trieste, Italy \\ [4.7cm]

\end{center}
\begin{center}
ABSTRACT
\end{center}

\vspace{3mm}

We present self-dual spin-3 and 4 actions using relevant Dreibein fields.
Since these actions start with a Chern-Simons like kinetic term (and
therefore)  cannot be obtained through dimensional reduction) one might wonder
whether  they need the presence of auxiliary ghost-killings fields. It turns
out that  they must contain, also in this three dimensional case, auxiliary
fields.  Auxiliary scalars do not break self-duality: their free actions does
not  contain kinetic terms.

\newpage
Self-dual theories for odd dimensions were discovered time ago by Townsend,
Pilch and van Nieuwenhuizen [1]. For abelian vector theories, they can be
shown to be classically and quantum mechanically equivalent [2] to the
Maxwell-Chern-Simons (MCS) [3] model, if one permits a non minimal coupling in
the  self-dual model while keeps the minimal one for the gauge invariant
second  order MCS theory.

Or one can assume minimal coupling in both cases and then, although both
models propagates one massive-spin 1 mode these theories will not be
equivalent.

Spin-2 presents a new feature: there are three topological spin-2 theories:
linearized topological massive gravity [4], a second order Einstein-CS
action [5] and the first order self-dual one [6]. In the vector case the
topological massive action is second order, whereas the self-dual one is
first order. Spin-two fields presents a new feature: exact topological
massive gravity [4] is a third order action while self-dual gravity [5] is,
by definition, first order. Self-dual gravity is a good example of the
relevance of the Dreibein representation [7] for higher spin gauge fields:
its more compact form is obtained when the spin two field is represented by
the (linearized) unsymmetrized second rank tensor $w_{pa}$ where $p$ is the
gauge index and $a$ is the flat remanent of a Lorentz index. Its gauge
variation is given by $\delta w_{pa}=\partial_p\xi_a$.

When dealing with higher spin particles ($s\geq 3$) one is always concerned
with whether they can have consistent interactions with either other basic
elementary systems or (at least) with themselves. Along this direction,
recently it has been shown the existence of higher-spin self interacting
bosonic theories [14]. These theories are third order in the basic fields,
their structure is very  similar to metric topological Chern-Simons gravity
[4].

In $d=4$, bosons obey second order field equations. Precisely due to this
fact, coupling them to abelian vectors (when charged) or to gravity (which
is always mandatory because of the universality of gravity) leads to consider
a  wide variety of different types of non minimal coupling, once the canonical
ones are shown not to work, as it is in general the case. The natural
solution to this problem comes from charged string theory models which
consistently contain in their spectrum all spins [15].

In dimension 3 we have the peculiarity of the existence of these first order,
Dirac-like, bosonic self-dual theories for spin 1 and 2. It seems to us
worthwhile to  construct flat models for spin 3 and 4 in order to investigate
whether they  can be consistently coupled to abelian vectors or to gravity.

Here we report about the precise, Dirac like, self-dual actions we found for
spin 3  and 4. We want to mention an additional (more technical) problem.

Massive spin-3 in dimensions $d\geq 4$ cannot avoid the presence of auxiliary
fields as it is clearly shown by dimensional reduction from its massless,
gauge invariant $d+1$ dimensional spin-3 ascendent action [8]. In $d=3$ it
is hard to imagine what might be the 4-dimensional ascendent of a three
dimensional self-dual action (whose kinetic term is essentially given by
$\sim w_{(3)}\epsilon \partial w_{(3)}$). Therefore, one might ask again
whether
self-dual pure spin-3 (or higher) needs the presence of auxiliary fields.
Even if self-dual spin-3 would not have needed auxiliary fields one should
ask what is the fate of spin-4 since the real high spin field is spin-4. This
is due to the fact, if one works in the symmetric representation where
$w_{(4)}$  is the basic 4-index symmetric tensor which carries the physical
massless excitations, $w_{(4)}$ has to be double traceless [9], i.e.,
$w\equiv w_{pprr}=0$. This condition is uniformly obeyed by any spins-s
grater than 4, v.e. $w_{pprr\ell_1\cdots \ell_{s-4}}=0$.

In the following we will show that both self-dual spin-3 and 4 actions
require the presence of self-dual auxiliary fields of spin-1 and 0 for the
former and spin-2 and 1 for the latter.

The symmetric formulation of massless spin-3 in $d\geq 3$ was given in [9].
The  first order Vierbein formulation was presented by Vasiliev [7] and a
second  order action was introduced in [10]. The associated massive spin-3
models are  discussed in [8].

In three dimensions there exist additional possibilities, (at the abelian
level) which perhaps, taking into account the analysis performed in [5] for
the spin-2 case, will be 3: the topological massive third order formulation
discovered by Damour and Deser [10], the first order self-dual action which
is presented here and the intermediate second order action equivalent to
these two similar to the spin-2 intermediate [12]. Since spin-3 is simpler
we treat if first.

Self-dual spin-3 action is the addition of three layers:
\begin{equation}
S=S_3+S_{31}+S_{10}
\end{equation}
were
\begin{eqnarray}
S_{3} &\equiv & 2^{-1}\mu <w_{p\bar{a}_1\bar{a}_2}\varepsilon^{pmn}
\partial_m w_{n\bar{a}_1\bar{a}_2}>-6^{-1}\mu^2<\varepsilon^{pmn}
\varepsilon^{abc}\eta_{pa}w_{m\bar{b}\bar{d}}w_{n\bar{c}\bar{d}}>,\\
S_{31} &\equiv & \mu^2<w_p u_p>+2^{-1}\alpha\mu <u_p\varepsilon^{pmn}
\partial_m u_n>+2^{-1}\beta\mu^2<u_pu_p>,\\
S_{10} &\equiv & \mu<\phi\partial_pu_p>+2^{-1}\gamma <\phi \square \phi >
+2^{-1}\delta\mu^2<\phi^2>.
\end{eqnarray}

In three dimensions $[\phi ]=m^{1/2}=[w]=[u]$. The basic field
$w_{p\bar{a}_1\bar{a}_2}$ is symmetric and traceless in its Dreibein Lorentz
indices
$w_{p\bar{a}_1\bar{a}_2}=w_{p\bar{a}_2\bar{a}_1},w_{p\bar{a}\bar{a}}=0$  while
$p$ is a world index, unrelated to them. (In the following, a set of  barred
indices will indicate that the associated tensor is symmetric and  traceless
in this set.) The algebraically irreducible descomposition of
$w_{p\bar{a_1}\bar{a_2}}$ is

$$
w_{p\bar{a}_1\bar{a}_2}=w_{\bar{p}\bar{a}_2\bar{a}_1}+
\varepsilon_{pa_1b}h_{\bar{b}\bar{a}_2}+\varepsilon_{pa_2b}
h_{\bar{b}\bar{a}_1}+b(\eta_{pa_1}w_{a_2}\eta_{pa_2}w_{a_1}-2(3)^{-1}
\eta_{a_1a_2}w_p). \eqno (5a)
$$

The 15 independent components of $w_{p\bar{a}_1\bar{a}_2}$ are represented by
the 7 components of $w_{p\bar{a}_1\bar{a}_2}$ plus the 5 needed to describe
$h_{\bar{b}\bar{c}}$ plus the last 3 which determine
$w_p\equiv w_{r\bar{r}\bar{p}}$, the unique nonvanishing trace of
$w_{p\bar{a}_1\bar{a}_2}$. Taking the trace in Eq.(5a) one obtains $b=3/10$
and  calculating the symmetric part of $\epsilon_b{}^{pa}w_{p\bar{a}\bar{a}}$
one  is led to determine $h_{\bar{b}\bar{c}}$:
$$
h=h_{\bar{b}\bar{c}}=-6^{-1}(\varepsilon_b{}^{pa}w_{p\bar{a}\bar{c}}+
\varepsilon_c{}^{pa}w_{p\bar{a}\bar{b}}).\eqno (5b)
$$

The first interesting fact is that $S_3$ has the good spin-3 and spin-2
behaviour. The associated field equations
$E^{p\bar{a}_1\bar{a}_2}\equiv \delta S^3/\delta w_{p\bar{a}_1\bar{a}_2}=0$
propagate one parity sensitive spin-3 excitation, do not propagate neither
the other possible spin-3 variable nor any spin-2 degree of freedom (those
contained in $h^T_{\bar{a}\bar{b}}$, the transverse part of
$h_{\bar{a}\bar{b}}:\partial_{\bar{a}}h^T_{\bar{a}\bar{b}}=0$. However, $S_3$
has spin-1 ghosts and this is the reason one has to add a second layer which
will fix this situation. $S_{31}$ is a pure self-dual vector action for the
auxiliary vector $u_p$  plus the simplest, contact term $\sim
<w\  \ u_p>$. In  general one might also consider terms
$\sim \mu <w_p\epsilon^{pmn}\partial_mu_n>$ but we have been lucky and
there is no need to include them. Addition of these two layers leads to
$S_3+S_{31}$ whose field equations are
\setcounter{equation}{5}
\begin{eqnarray}
E^{p\bar{a}_1\bar{a}_2} &\equiv & \varepsilon^{pmn}\partial_m
w_{n\bar{a}_1\bar{a}_2}+6^{-1}\mu (\eta_{pa_1}w_{a_2}+\eta_{pa_2}w_{a_1}-
w_{a_1\bar{p}\bar{a}_2}w_{a_2\bar{p}\bar{a}_1})\nonumber \\
& & +2^{-1}\mu (\eta_{pa_1}u_{a_2}+\eta_{pa_2}u_{a_1}-2(3)^{-1}\eta_{a_1a_2}
u_p)=0,\\
F^p &\equiv & \alpha\varepsilon^{pmn}\partial_mu_n +\beta\mu u_p +\mu
w_p=0.
\end{eqnarray}

These two equations can be analyzed by further breaking of the algebraic
decomposition (5a) in terms of its $SL(2,R)$ irreducible representations. We
introduce the three dimensional covariant (and non local) $T$-projectors
which, in the vector case, are
$$
u_p=u^T_p+\wh{\partial}_pu^L,\  \  \ \wh{\partial}_p\equiv \square^{-1/2}
\partial_p,
$$
$$
\wh{\partial}_pu^T_p=0,\  \  \ \wh{\partial}_p\cdot \wh{\partial}_p=1.
\eqno (8a)
$$

For spin-2 and 3, similar decompositions for symmetric traceless second and
third rank tensors have the form:
$$
h_{\bar{p}\bar{a}}=h^T_{\bar{p}\bar{a}}+
\wh{\partial}_{(\bar{p}}h^L_{\bar{a})},\  \  \
\wh{\partial}_ph^T_{\bar{p}\bar{a}}=0=h^T_{\bar{p}\bar{p}},\eqno (8b)
$$
$$
w_{\bar{p}\bar{a}\bar{b}}=w^T_{\bar{p}\bar{a}\bar{b}}+
\wh{\partial}_{(p}w^L_{\bar{a}\bar{b})},\  \  \
\wh{\partial}_pw^T_{\bar{p}\bar{a}\bar{b}}=0=w^T_{\bar{p}\bar{p}\bar{b}}.
\eqno (8c)
$$

Symmetric traceless transverse $3d$ tensors
$(u_p^T,h^T_{\bar{p}\bar{a}},w^T_{\bar{p}\bar{a}\bar{b}}w^T_{\bar{p}\bar{a}
\bar{b}\bar{c}})$ have two independent components corresponding to the two
$P$-sensitive pseudospin-$j(j=1,2,3,4)$ excitation they can propagate. A
final covariant spliting of these set (symmetric, traceless, transverse)
tensors is obtained by means of the pure pseudospin-$j$ projectors
$p_j^\pm w^T_{\bar{p}\bar{a}\bar{b}\cdots \bar{c}}$ [6]
\setcounter{equation}{8}
\begin{equation}
p^\pm_jw^T_{\bar{p}\bar{a}\bar{b}\cdots \bar{c}}\equiv
w^{T^\pm}_{\bar{p}\bar{a}\bar{b}\cdots \bar{c}}=2^{-1}
w^T_{\bar{p}\bar{a}\bar{b}\cdots \bar{c}}\pm\frac{1}{2j}
\varepsilon_{(p}{}^{mn}\wh{\partial}_m
w_{\bar{n}\bar{a}\bar{b}\cdots \bar{c})},
\end{equation}
where the indicated symmetrization is the minimal one and does not carry a
normalization coefficient. It is straightforward to check that
\begin{equation}
p^+_j+p_j^-={\bf 1},\  \ p_j^+-p_j^-=\frac{1}{j}
\varepsilon (.\ddot{\   }\wh{\partial}\cdots ).
\end{equation}

Armed with these projectors one can analyse the behaviour of
$E^{\bar{p}\bar{a}\bar{b}T}$, the spin-3 sector of Eq.(6). It turns out that
$E^{\bar{p}\bar{a}\bar{b}T}$ propagates the spin-3$^+$ part of
$w^T_{\bar{p}\bar{a}\bar{b}}$ and annihilates
$w^{T-}_{\bar{p}\bar{a}\bar{b}}$. Then ones goes to the spin-2 sector and it
is immediate to verify that
$\partial_p E^{p\bar{a}b},\check{E}^{\bar{b}\bar{c}}\equiv
\varepsilon_{(bpa}E^{p\bar{a}}{}_{\bar{c})}$ do not allow the propagation of
$h^{T^\pm}_{\bar{a}\bar{b}}$. The spin-1 dynamical behaviour is determined
by $\partial_{pa}E^{p\bar{a}\bar{b}}$,
$\partial_b\check{E}^{\bar{b}\bar{a}},E^b\equiv E^{p\bar{p}\bar{b}}$ and
$F^p$. In order not to have any spin-1 excitation alive we must choose

\begin{equation}
\alpha = \beta =-18.
\end{equation}

Unfortunately this is not the last step in order to get a pure pseudospin-3$^+$
propagation. $S_3+S_{31}$ has scalar ghosts and therefore they have to be
destroyed by an auxiliary scalar $\phi$.

This is the reason of having to add to the first two layers $S_3+S_{31}$ the
last one, $S_{10}$ defined in Eq.(4). In principle one should have to
consider the posibility of kinetic terms like $\sim \phi \square \phi$ which
are the second order and therefore would break the full system self-duality.
The fields equations derived from $S$ are

\begin{eqnarray}
\delta_w S &\sim & E^{p\bar{a}_1\bar{a}_2} = 0 \\
\delta_{u} S &\sim & 'F^p \equiv F^p -\partial_p \phi = 0 , \\
\delta_\phi S &\sim & G\equiv \gamma \square \phi +\delta \mu^2\phi +
\mu\partial_p u_p = 0.
\end{eqnarray}

There are five scalar excitations which the system might propagate
$\wh{\partial}_{pab}w_{\bar{p}\bar{a}\bar{b}},\wh{\partial}_{ab}
h_{\bar{a}\bar{b}}$, $\wh{\partial}_pw_p$, $\wh{\partial}_p u_p$,
$\phi$. However, since $\partial_p E^{p\bar{a}\bar{b}}$ and
$\check{E}^{\bar{b}\bar{c}}$ tells us that
$$
\mu h_{\bar{b}\bar{c}}=-3(\partial_bu_c+\partial_cu_b-2(3)^{-1}\eta_{ab}
(\partial \cdot u)),\eqno (15a)
$$
$$
\partial_bw_c+\partial_cw_b-(\partial_pw_{b\bar{p}\bar{c}}+\partial_p
w_{c\bar{p}\bar{b}})+3(\partial_bu_c+\partial_cu_b-2(3)^{-1}\eta_{bc}
(\partial \cdot u))=0,\eqno (15b)
$$
it is immediate that, if neither $\wh{\partial}_p u_p$ nor
$\wh{\partial}_p w_p$ propagate (i.e., $\wh{\partial}_p u_p = 0 =
\wh{\partial}_p w_p)\wh{\partial}_{pab}w_{\bar{p}\bar{a}\bar{b}}$
and $\wh{\partial}_{pa}h_{\bar{p}\bar{a}}$ will not propagate either.
The key equations are the vanishing of $\partial_b E^{p\bar{p}\bar{b}}$,
$\partial_p `F^p$ and $G$ where in the first one, makes use of Eqs.(5a) and
(15).

They can be written, respectively
$$
(12 \square +5(8)^{-1}\mu^2)\partial \cdot u +2^{-1}\mu^2\partial \cdot w=0,
\eqno (16a)
$$
$$
\mu\beta\partial \cdot u+\mu\partial \cdot w -\square \phi =0, \eqno (16b)
$$
$$
\mu\partial \cdot u +(\gamma \square +\delta \mu^2)\phi=0. \eqno (16c)
$$

Introducing the dimensionless operator $x\equiv \mu^{-1}\square^{1/2}$ it is
straightforward to see that the inverse propagator of
$\wh{\partial}\cdot w,\wh{\partial}\cdot u, \phi$ is
\setcounter{equation}{16}
\begin{equation}
\Delta (x)\equiv -(\gamma x^2 +\delta )(12x^2+5(8)^{-1})+
2^{-1}x^2+2^{-1}\beta (\gamma x^2+\delta ).
\end{equation}

These scalar variables (and consequently
$\wh{\partial}_{pab}w_{\bar{p}\bar{a}\bar{b}},\wh{\partial}_{pa}
h_{\bar{p}\bar{a}}$) do not propagate if the polynomial $\Delta (x)$ becomes
zero order, i.e., $\Delta (x)\equiv \Delta_4\cdot x^0=\Delta_4\cdot 1$. This
condition uniquely determines $\gamma ,\delta$
\begin{equation}
\gamma =0,\  \  \ \delta =(24)^{-1}.
\end{equation}

Note that the vanishing of $\gamma$ makes action $S_{10}$ first order
(scalars appear of the self-dual type too), leading to the final $S$ being
fully first order. Observe that we do not claim mathematical uniqueness for a
pure spin-3$^+$ (or 3$^+$) $3d$ action: in the scalar sector one could have
consider coupling terms like $\sim\phi (\partial \cdot w)$. However, it seems
to us that, if one starts with the right-spin Dreibein seed (in the case
$S_3$), then $S_{31}$ is unique if we demand that it must be the vector
self-dual action coupled in the softest possible ways to $S_3$ (the coupling
term must be, at most, first order and if possible algebraic). The
construction of the auxiliary scalar action $S_{10}$ again is unique: it
contains the free self-dual scalar action ($\sim \mu^2\phi^2$, no
Klein-Gordon kinetic term) and it is next-neighbour coupled to the auxiliary
spin-1 field, discarding $\phi (\partial \cdot w)$ which is not of the
next-neighbour type.

All these results will be useful when dealing with the much complex case of
spin-4.

We start this analysis by introducing the spin-4 part of the final action
$S_{42}$  with the right physical behaviour up to the spin-2 sector. It reads
\begin{eqnarray}
S_{42} &\equiv & (2)^{-1}\mu <w_{p\bar{a}\bar{b}\bar{c}}\varepsilon^{pmn}
\partial_m w_{n\bar{a}\bar{b}\bar{c}}>-2^{-1}\mu^2 <\varepsilon^{pmn}
\varepsilon^{abc}\eta_{pa}w_{m\bar{b}\bar{d}_1\bar{d}_2}
w_{n\bar{c}\bar{d}_1\bar{d}_2}>\nonumber \\ [3mm]
& & +\mu^2<w_{p\bar{p}\bar{a}\bar{b}}u_{ab}>+(2)^{-1}\alpha\mu <u_{pa}
\varepsilon^{pmn}\partial_m u_{na}>+2^{-1}\beta\mu^2<\varepsilon^{pmn}
\varepsilon^{abc}\eta_{pa}u_{mb}u_{nc}>,
\end{eqnarray}
where $w_{p\bar{a}\bar{b}\bar{c}}$ is symmetric and traceless (ST) in its
three last barred indices and $u_{pa}$ is an auxiliary self-dual second rank
tensor, $[w]=[u]=m^{1/2}$. Their algebraically irreducible representations
are, respectively
\begin{equation}
w_{p\bar{a}\bar{b}\bar{c}}=w_{\bar{p}\bar{a}\bar{b}\bar{c}}+
\varepsilon_{p(ad}h_{d\bar{b}\bar{c})}+5(21)^{-1}\eta_{p(a}w_{\bar{b}\bar{c})}
-2(21)^{-1}w_{p(\bar{a}}\eta_{bc)},
\end{equation}
$$
u_{pa}=u_{\bar{p}\bar{a}}+\varepsilon_{pad}h_d+3^{-1}\eta_{pa}u,\  \
h_d=-2^{-1}\varepsilon_d{}^{pa}u_{pa},\eqno (21a,b)
$$
where $w_{p\bar{p}\bar{b}\bar{c}}\equiv w_{\bar{b}\bar{c}}$ and $u_{pp}$ are
the unique non-vanishing contractions which can be made out of
$w_{p\bar{a}\bar{b}\bar{c}}$ and $u_{pa}$, respectively. Symmetrizations are
minimal with coefficient one in front and sets of barred indices continue to
indicate ST tensors.

Variations with respect the $w_{p\bar{a}\bar{b}\bar{c}}$ and $u_{pa}$ yield
the initial set of field equations
\setcounter{equation}{21}
\begin{eqnarray}
E_{p\bar{a}\bar{b}\bar{c}} & \equiv & \varepsilon_p{}^{mn}\partial_m
w_{n\bar{a}\bar{b}\bar{c}}+\mu (3)^{-1}\{\eta_{p(a}w_{\bar{b}\bar{c})}-
w_{(a\bar{b}\bar{c})\bar{p}}\}\nonumber\\
& & +\mu(3)^{-1}\{\eta_{p(a}u_{\bar{b}\bar{c})}-2(5)^{-1}
\eta_{(ab}u_{\bar{c}\bar{p})}\}=0,\\
F_{pa}&\equiv &\mu w_{\bar{p}\bar{a}}+\alpha \varepsilon_p{}^{mn}\partial_m
u_{na}+\mu\beta \varepsilon_p{}^{mn}\varepsilon_a{}^{bc}\eta_{nc}u_{mb}=0.
\end{eqnarray}

The spin-4$^\pm$ excitations are carried on the transverse part of
$w_{\bar{p}\bar{a}\bar{b}\bar{c}}:w^T_{\bar{p}\bar{a}\bar{b}\bar{c}},
\partial_p w^T_{\bar{p}\bar{a}\bar{b}\bar{c}}=0$ while there are two sets of
spin-3 variables: those contained in
$\wh{\partial}_p w_{\bar{p}\bar{a}\bar{b}\bar{c}}$ and those defined by
$h^T_{\bar{a}\bar{b}\bar{c}}$. Use of the spin-4$^\pm$projectors defined in
Eqs.(9) and (10) show that $E_{p\bar{a}\bar{b}\bar{c}}$ uniquely propagate
spin-4$^+$ (make the spin-4$^-$ degree of freedom to cancel) and does not
propagate neither ($\wh{\partial}_pw_{\bar{p}\bar{a}\bar{b}\bar{c}})^T$ nor
$h^T_{\bar{a}\bar{b}\bar{c}}$. In fact, equations
$\partial_pE_{p\bar{a}\bar{b}\bar{c}}=0=\varepsilon_{(a}{}^{pd}
E_{p\bar{d}\bar{b}\bar{c})}$ are equivalent to
\begin{eqnarray}
4\mu h_{\bar{a}\bar{b}\bar{c}} &=& 2(5)^{-1}\eta_{(ab}\partial_p
u_{\bar{p}\bar{c})}-\partial_{(a}u_{\bar{b}\bar{c})},\\
\partial_{(a}w_{\bar{b}\bar{c})}-\partial_{p}w_{(a\  \bar{p}\bar{b}\bar{c})}
&=&
2(5)^{-1}\eta_{(ab}\partial_pu_{\bar{p}\bar{c})}-\partial_{(a}u_{\bar{b}\bar{c})}.
\end{eqnarray}

These equations say both $h_{\bar{a}\bar{b}\bar{c}}$ and $\partial_p$
$w_{\bar{p}\bar{a}\bar{b}\bar{c}}$ are curls of spin-2 objects and therefore
their pure spin-3 parts have to vanish.

Four variables describe the spin-2 sector of
$S_{42}:(\wh{\partial}_{pa}w_{\bar{p}\bar{a}\bar{b}\bar{c}})^T$,
$(\wh{\partial}_ph_{\bar{p}\bar{a}\bar{b}})^T$, $w^T_{\bar{p}\bar{a}}$,
$u^T_{\bar{p}\bar{a}}$. The equations which determine their dynamical
behaviour are $\partial_{pa}E^{p\bar{a}\bar{b}\bar{c}}=0$,
$\check{E}^{\bar{a}\bar{b}\bar{c}}=0$,
$E_{\bar{b}\bar{c}}\equiv E_{p\bar{p}\bar{b}\bar{c}}=0$ and $F_{pa}=0$. After
some algebra one is led to a separated propagation eqaution for
$u^T_{\bar{p}\bar{a}}\equiv \omega ,p^\pm \omega \equiv \omega^\pm$
\begin{eqnarray}
(x^2 &+& 7(5)^{-1}-4(3)^{-1}\beta )(\omega^++\omega^-)+2(3)^{-1}x(\alpha
x+\beta ) \omega^++2(3)^{-1}x(\alpha x -\beta )\omega ^-\nonumber \\
& & -4(3)^{-1}\alpha x(\omega^+-\omega^-)=0.
\end{eqnarray}

Projecting on this spin-2$^+$ (2$^-$) subspaces we obtain the two uncoupled
equations which determine their evolution
\begin{equation}
\{ x^2(1+2(3)^{-1}\alpha )\mp 2(3)^{-1}(2\alpha -\beta )x+(7(5)^{-1}-4(3)^{-1}
\beta )\}\omega^\pm =0,
\end{equation}
(either all upper indices or all down right). Non-propagations of one of
these two variables determines the values of $\alpha\beta$:
\begin{equation}
\alpha = -3(2)^{-1},\  \ \beta =-3,
\end{equation}
and, due to Eq.(27), entails the non-propagation of the other companion
variable. $S_{42}$ (19) has been uniquely determined requesting its good
physical behaviour in its highest spin sector ($s=4,3,2$).

However, it contains vector and scalar ghosts. This is the reason why we have
to add two additional layers. The most difficult of them is spin-1 fixing
action. Its ambiguity stems in the wide range of mathematically consistent
terms one might have to consider $ab$ initio.

In principle $S_{21}$ may be
\begin{eqnarray}
S_{21} &\equiv & -2\lambda_1\mu <h_a\partial_bu_{\bar{a}\bar{c}}>+2\lambda_2
\mu <v_p\partial_ru_{\bar{r}\bar{p}}>\nonumber \\
&  & \gamma_2\mu <h_a\varepsilon^{abc}\partial_bh_c>+\gamma_1 (2)^{-1}\mu
<v_p \varepsilon^{pmn}\partial_mv_n>\nonumber \\
& & +\rho \mu^2<h^2_a>+\delta (2)^{-1}\mu^2<v_a^2>+2\varepsilon \mu^2 <h_pv_p>
+2\kappa \mu <h_a \partial_b w_{\bar{b}\bar{a}}>\nonumber \\
& & +2\varphi \mu <v_p\partial_r w_{\bar{r}\bar{p}}>+2\sigma \mu
<v_p \varepsilon^{pmn}\partial_m h_n>,
\end{eqnarray}
which can be regarded as the addition of the self-dual action for the spin-1
variable $h_a$ contained in $u_{pa}$ plus the auxiliary self-dual action for
the auxilary vector $u_p$ algebraically coupled through $\sim h\cdot v$ plus
more bizarre  terms like $\sim h_a \partial_b u_{\bar{b}\bar{a}}$,
$h_a\partial_b w_{\bar{b}\bar{a}}$, $v_a\partial_b u_{\bar{b}\bar{a}}$,
$v_a \varepsilon^{abc}\partial_b h_c$ and the exotic term
$\sim v_a\partial_b w_{\bar{b}\bar{a}}$. We will not consider them, the
first because we already have chosen a good kinetic term for $u_{pa}$
($u_{pa} \varepsilon^{pmn}\partial_m u_{na}$ as in Eq.(19)), the last one
because it is not of the next-neighbour type (it is spin-4$\cdot$spin-1) and
second, third and fourth because we have decided to choose, whenever possible,
algebraic couplings and we have already a spin-2$\cdot$spin-1 contact term
$\sim h.v$. Therefore we rule out the present of terms  $\sim v_a \partial_b
u_{\bar{b}\bar{a}}v_a\varepsilon^{abc}\partial_bh_c$ as  well as the need form
a term $\sim h_a \partial_b w_{\bar{b}\bar{a}}$, a  different coupling term
linking spin-4 with spin-2 for the same reason. In  other words we take
$\lambda_1=\lambda_2=\kappa =\sigma =\varphi =0$ in  $S_{21}$.

Taking into account Eq.(21b) we write down in the modified spin-2 field
equations which govern this system (note that
$E^{\bar{p}\bar{a}\bar{b}\bar{c}}=0$ remains intact). They have the aspect
\begin{equation}
`F_{pa}\equiv F_{pa}+\gamma_2(\partial_ph_a-\partial_a h_p)-\rho
\varepsilon_{pab}h_b-\varepsilon \varepsilon_{pab} v_b =0.
\end{equation}

An additional vector-like field equation appears after varying $v_p$,
\begin{equation}
G_p\equiv \gamma_1 \varepsilon_p^{mn}\partial_mv_n +\delta\mu v_p+2
\varepsilon\mu h_p =0.
\end{equation}

We want to determine $\gamma_1,\gamma_2,\rho ,\delta ,\varepsilon$ in such
a way that none of the six spin-1 variables:
$\omega_8\equiv (\wh{\partial}_{pab}w_{\bar{p}\bar{a}\bar{b}\bar{c}})^T$,
$\omega_9\equiv (\wh{\partial}_{pa}h_{\bar{p}\bar{a}\bar{b}})^T$,
$\omega_{11}\equiv (\wh{\partial}_p u_{\bar{p}\bar{a}})^T$,
$\omega_{11}\equiv h^T_p$, $\omega_{12}\equiv h^T_p$, $\omega_{13}\equiv
v^T_p$ can propagate. Since  $\omega_8$ is given by
$\partial_{pab}E_{p\bar{a}\bar{b}\bar{c}}$ in terms of  the five remaining
variables $\omega_9\cdots \omega_{13}$ we go after the non  propagation of
them.

They are determined by $\partial_{ab}\check{E}_{\bar{b}\bar{a}\bar{c}}=0$,
$\partial_bE_{\bar{b}\bar{c}}=0$, $\partial_p `F_{pa}=0$,
$`\check{F}^b=0$ and $G^p=0$ After minor algebra and some use of Eq.(24) the
five equations become
\begin{eqnarray}
& &4\mu\partial_{ab}h_{\bar{a}\bar{b}\bar{c}}+8(5)^{-1}\square \partial_a
u_{\bar{a}\bar{c}}+5^{-1}\partial_c(\partial_{ab}u_{\bar{a}\bar{b}})=0,\\[3mm]
&
&-4\partial_{ab}h_{\bar{a}\bar{b}\bar{c}}-3^{-1}\varepsilon_c{}^{pr}\partial_p
(\partial_bw_{\bar{b}\bar{r}}+4(3)^{-1}\mu\partial_pw_{\bar{p}\bar{c}}+
7(5)^{-1}\mu\partial_p u_{\bar{p}\bar{c}}=0,\\[3mm]
& &\mu\partial_pw_{\bar{p}\bar{a}}-3\mu\partial_pu_{\bar{p}\bar{a}}+(\rho -3)
\mu\varepsilon_a{}^{pr}\partial_ph_r+2\mu\partial_au+\nonumber \\
& & \  \  \  \  \ +\gamma_2(\square
h_a-\partial_a(\partial_ph_p))+\varepsilon\mu\varepsilon_a
{}^{pr}\partial_pv_r=0,\\[3mm]
 & &3(2)^{-1}\partial_pu_{\bar{p}\bar{b}}+2(\rho -3)\mu h_b +2\varepsilon\mu
v_b+ (2\gamma_2+2(3)^{-1})\varepsilon_b{}^{pr}\partial_ph_r-
\partial_bu=0
\end{eqnarray}
and Eq.(31) as it stands.

Working in a similar way to what we did for the spin-3 case, the vanishing of
$\omega_9\cdots \omega_{13}$ is equivalent to their non propagation and this
is reached if $\Delta (x)=\Delta_0x^4+\cdots +\Delta_4\cdot 1$ becomes
$\Delta_4 \cdot 1$. Straighforward calculations give
\begin{eqnarray}
\Delta
(x)=&-&3(10)^{-1}\gamma_1(9\gamma_2+8)x^4+\{3(2)^{-1}\gamma_1(1-9(5)^{-1}
\rho ')-3(5)^{-1}\delta (9(2)^{-1}\gamma_2+4)\}x^3\nonumber \\
&+&\{-27(5)^{-1}\gamma_1(2\gamma_2+3(2)^{-1})-27(10)^{-1}\delta\rho '+
3(2)^{-1}\delta +27(5)^{-1}\varepsilon^2\}x^2\nonumber \\
&-&27(5)^{-1}\{\delta (2\gamma_2+3(2)^{-1})+2\gamma_1\rho '\}x +54(5)^{-1}
\{2\varepsilon^2-\delta \rho '\}\cdot 1,
\end{eqnarray}
where for convenience $\rho '\equiv \rho -3$. Requesting the vanishing of the
coefficients $\Delta_{0,1,2,3}$ of the inverse propagator $\Delta_x$ one is
lead to
\begin{eqnarray}
& &\gamma_2=-8(9)^{-1},\  \  \rho '=5(9)^{-1}=\rho -3,(\rho
=-4\gamma_2),\nonumber \\
& &\gamma_1=-18(5)^{-1}\varepsilon^2,\  \  \delta =4\gamma_1=-72(5)^{-1}
\varepsilon^2.
\end{eqnarray}

Redefining 2 $\varepsilon v_p \to v_p$ the final unique form of $S_{21}$
becomes
$$S_{21}=-8(9)^{-1}\mu<h_a\varepsilon^{abc}\partial_bh_c>-9(20)^{-1}\mu
<v_p\varepsilon^{pmn}\partial_mv_n>
$$
$$
+32(9)^{-1}\mu^2<h_a^2>-9(5)^{-1}\mu^2<v^2_p>+<h_pv_p>\mu^2.\eqno (29b)
$$

The action $S_{42}+S_{21}$ has the right physical properties up to spin-1.
However, its scalar sector contains ghost which we have to exorcise by
introducing an auxiliary self-dual scalar $\phi$. Its associated action
$S_{10}$ constitutes the last layer we need to determine the final pure
self-dual spin-4$^+$ action S.

The most general scalar auxiliary action one can add to $S_{42}+S_{21}$ is
\setcounter{equation}{37}
\begin{eqnarray}
S_{10}&\equiv &2a_1\mu
<\phi\partial_pu_p>+2a_2\mu <\phi\partial_ph_p>+2a_7\mu
<u\partial_ph_p>+2a_8\mu <u\partial_pv_p>\nonumber \\
& & a_5\mu^2 <\phi
u>+2^{-1}a_3\mu^2<\phi^2>+2^{-1}a_4<\phi \square \phi>\nonumber \\
& &
+2^{-1}a_6\mu^2<u^2>+2^{-1}a_9<u\square u>+a_{10}<u\square \phi
>.
\end{eqnarray}

Taking advantage of what we learned from the spin-3 case, we assume that
there will be a final scalar auxiliary action fully self-dual, i.e., that
there exists a non trivial $S_{10}$ with vanishing $a_4,a_9$ and $a_{10}$. We
also assume a vanishing $a_7$, since this term can be seen as an unpleasant
kinetic term to add to the self-dual actions
$u_{\bar{p}\bar{a}}\varepsilon^{pmn}\partial_m u_{\bar{n}\bar{a}}$ and
$h_p\varepsilon^{pmn}\partial_m h_n$. The final equations are
$$
E_{p\bar{a}\bar{b}\bar{c}}=0,\eqno (22)
$$

\vspace{-5mm}
\setcounter{equation}{38}
\begin{eqnarray}
``F_{pa}&\equiv &`F_{pa}+\mu a_5\eta_{pa}\phi +a_2\varepsilon_{pa}
{}^m\partial_m\phi +a_6\mu\eta_{pa}u +2a_8\eta_{pa}(\partial \cdot
v)=0,\\[3mm]
`G_p&\equiv &G_p-2\alpha_1\mu\partial_p\phi-2a_8\partial_p u=0,\\[3mm]
H&\equiv &\delta S_{10}/\delta\phi =2a_1(\partial \cdot v)+2a_2(\partial \cdot
h)+a_4\mu u+\mu a_3\phi=0.
\end{eqnarray}

The scalar sector has eight independent variables:
\begin{eqnarray}
\omega_1&\equiv &\wh{\partial}_{pabc}w_{\bar{p}\bar{a}\bar{b}\bar{c}},\  \
\omega_2\equiv \wh{\partial}_{pab}h_{\bar{p}\bar{a}\bar{b}},\  \
\omega_3\equiv \wh{\partial}_{ab}w_{\bar{a}\bar{b}},\nonumber \\
\omega_4 &\equiv &\wh{\partial}_{ab}u_{\bar{a}\bar{b}},\  \
\omega_5\equiv \wh{\partial}_ah_a,\omega_6\equiv \mu u,\nonumber \\
\omega_7&\equiv &\wh{\partial}_av_a,\  \ \omega_8\equiv \mu \phi
\end{eqnarray}
whose evolution is determined by
$\partial_{pabc}E_{p\bar{a}\bar{b}\bar{c}},\partial_{abc}
\check{E}_{\bar{a}\bar{b}\bar{c}},\partial_{bc}E_{\bar{b}\bar{c}},
\partial_{pa}``F_{pa},\partial_{b}``\check{F}_{b},\partial_{p}`G_{p}$ and
$H$.

The first set of 3 equations is derived from Eq.(22) taking into account the
algebraic structure of $w_{p\bar{a}\bar{b}\bar{c}}$ as given in Eq.(20). It
turns out to be
\begin{eqnarray}
& &-5\partial_{pabc}w_{\bar{p}\bar{a}\bar{b}\bar{c}}+5(21)^{-1}\square
\partial_{pa}w_{\bar{p}\bar{a}}+3\square \partial_{\bar{p}\bar{a}}
u_{\bar{p}\bar{a}}=0,\\
& &4\mu \partial_{pab}h_{\bar{p}\bar{a}\bar{b}}+9(5)^{-1}\mu
\partial_{pa}u_{\bar{p}\bar{a}}=0,\\
& &-4\partial_{pab}h_{\bar{p}\bar{a}\bar{b}}+4(3)^{-1}\mu\partial_{pa}
w_{\bar{p}\bar{a}}+7(5)^{-1}\mu\partial_{pa}u_{\bar{p}\bar{a}}=0.
\end{eqnarray}

The second set comes from Eq.(39). It consists of
\begin{eqnarray}
& &\partial_{pa}``F_{pa}\equiv \mu\partial_{pa}w_{\bar{p}\bar{a}}-3\mu
\partial_{pa}u_{\bar{p}\bar{a}}+\mu (2+a_6)\square u+\nonumber \\
& &+\mu a_5\square \phi +2a_8\square \partial_pv_p=0,\\[3mm]
& &\partial_b``\check{F}_b\equiv 3(2)^{-1}\partial_{pa}w_{\bar{p}\bar{a}}+
10(9)^{-1}\mu\partial_ph_p+\mu\partial_pv_p-\nonumber \\
& &-\square u-2a_2\mu \square \phi =0,\\[3mm]
& & ``F_{pp}\equiv \partial_ph_p+(2+a_6)\mu u+a_5\mu\phi +2a_8\partial_p
v_p=0.
\end{eqnarray}

The last two equations are
\begin{equation}
\partial_p `G_p\equiv \delta_\mu\partial_pv_p+\mu \partial_ph_p-2a_1\mu \square
\phi-2a_8\square u=0,
\end{equation}
and Eq.(41) $H=0$. In terms of the $\omega$-variables (42) Eqs.(43)-(45)
allow to obtain $\omega_1,\omega_2,\omega_3$ as a function of $\omega_4$. In
particular
\begin{equation}
\omega_3=-3(20)^{-1}(9x^2+7)\omega_4.
\end{equation}

Then it is immediate to realize that Eqs.(46)-(49), (41) become a decoupled
subset of the full system. It can be written as
\begin{eqnarray}
& &-27(20)^{-1}(x^2+3)\omega_4+(2+a_6)\omega_6+2a_8x\omega_7+a_5\omega_8=0,\\
& &3(2)^{-1}x\omega_4+10(9)^{-1}\omega_5+\omega_7-x\omega_6-2a_2x\omega_8=0,\\
& &x\omega_5+(2+a_6)\omega_6+2a_8x\omega_7 +a_5\omega_8=0,\\
& &\omega_5-2a_8x\omega_6+\delta\omega_7-2a_1x\omega_8=0,\\
& &2a_2x\omega_5+a_5\omega_6+2a_1x\omega_7+a_3\omega_8=0.
\end{eqnarray}

The inverse of this determinat $\Delta_{(a_1,a_2,a_3,a_5,a_6,a_8)}$ is the
system's propagator. We wish to determine the $a_1\cdots a_8$ coefficients in
such a way that $\Delta (x)$ is a non-vanishing real number. First we
investigate the possibility of having a solution with pure next-neighbours
coupling terms, i.e., where $a_2=0=a_5$ (they are spin-2-spin-0 couplings).

In this case
\begin{eqnarray}
& &\Delta (a_2=0=a_5)=-27(20)^{-1}x^2(x^2+3)(4a^2_1x^2+a_3(\delta -2a_8))
\nonumber \\
& &-18x^2(a^2_1a_6'+a_3a_8^2)-9(2)^{-1}\delta a_3a_6',
\end{eqnarray}
where $a_6'\equiv 2+a_6$. Vanishing of its highest power coefficient leads
to
$$
a_1=0,\eqno (57a)
$$
and subsequent cancellation of quartic and quadratic terms impose
$$
a_3=0, \eqno (57b)
$$
which seem an inconsistent possibility, since in this case $\Delta (56)$
becomes  identically zero. However, since we are now thinking of not having
$\phi$-dependent actions $(a_1=a_2=a_3=a_5=0)$ we have to consider the
appropiate system of field equations which consists of Eqs.(22),(39) and (40)
for these values of $a_{1,2,3,5}$ and does no longer contain Eq.(41). Its
crucial decoupled part consists of Eqs.(51)-(54) $(a_1=a_2=a_3=a_5=0)$ and
the non propagating character is determined by imposing to its associated
(quartic) determinant to be a non zero real number. This leads us to
determine $a_6$ and $a_8$
\setcounter{equation}{57}
\begin{equation}
a_6=(5)^{-1}44,\  \  \ a_8=-9(10)^{-1}.
\end{equation}
$S_{10}$ attains a very simple form
\begin{equation}
S_{10}=-9(5)^{-1}\mu <u\partial_p v_p>+22(5)^{-1}\mu^2<u^2>,
\end{equation}
where there is no auxiliary scalar field present

This is the minimal solution. If one relaxes a little bit the assumption
of considering only next-neighbours coupling and investigate the consequence
of only imposing $a_2=0$ (leaving room for an algebraic non-next-neighbour
spin-2-spin-0 coupling) we are led to $a_1=a_3=0,a_6,a_8$ arbitraries and
$a_5$ arbitrary non-vanishing.

Similarly, one might constraint $a_5$ to vanish and try to determine $a_2$.
In this case one obtains (after redefining $\phi \to a_2\phi )$
\begin{eqnarray}
& & a_1=2^{-1}\delta ,a_2=1, a_3=20\delta^2(2+a_6)(6a_6+12-5\delta^2)^{-1},
\nonumber \\
& &a_6\neq 44(5)^{-1},a_8=4^{-1}\delta ,
\end{eqnarray}
and the corresponding full action is a pure spin-4$^+$ action too.

It is worth observing that simplest, self-dual, next-neighbour coupled pure
spin-4$^+$ is then given by:
\begin{equation}
S=S_{42}(19)+S_{21}(29a)+S_{10}(59)
\end{equation}
and contains only one auxiliary self-dual spin-2, $u_{ra}$, and one
(self-dual)  vector auxiliary field $v_r$, in addition to the fundamental
physical spin-4  carrier $w_{r\bar{a}\bar{b}\bar{c}}$.

In conclusion we have been able to uniquely construct self-dual spin-3 and
4 actions where auxiliary fields also appear in a self-dual form (including
scalars) and where coupling terms are next-neighbours. In both cases we
needed one self-dual auxiliary filed of spin s-2, s-3, up to spin-1.

Since spin-4 clearly is the higher-spin case we may conjecture that this
self-dual picture exists for arbitrary integer spin, where the unique non
uniform structure is the final layer fixing the good spin-0 behaviour.

An additional interesting question is what should be the higher spin
structure of topologically massive theories. We are inclined to think that
all of them will be of third-order, as it is the case for gravity and spin-3.

It would also be interesting to see what is the connection between the
present self-dual spin-3, and 4 formulations and the recently proposed [13]
anyonic relativistic actions for spin-$j$ real, since this scheme
consistently contains the self-dual abelian vector case.

However, as we mentioned in the beginning, whether this Dirac-like bosonic
structures can be consistently coupled either to abelian vectors or to
gravity is a worthwhile question which deserves further analysis.

\vspace{3cm}
\begin{center}
{\bf Acknowledgments}
\end{center}
\vspace{3mm}

One of the authors (A.K.) would like to thank Professor Abdus Salam, the
International Atomic Energy Agency and UNESCO for hospitality at the
international Centre for Theoretical Physics, Trieste.

\newpage
\begin{center}
REFERENCES
\end{center}
\vspace{3mm}
\begin{enumerate}
\item{}P.K. Townsend, K. Pilch and P. van Nieuwenhuizen, Phys. Lett. {\bf 136B}
(1984) 38.
\item{}S. Deser and R. Jackiw, Phys. Lett. {\bf 139B} (1984) 371.
\item{}W. Siegel, Nucl. Phys. Lett. {\bf B156} (1979) 135;
\par R. Jackiw and S. Templeton, Phys. Rev. {\bf D2} (1981) 2291;
\par J. Sch\"onfeld, Nucl. Phys. {\bf B185} (1981) 157.
\item{}S. Deser, R. Jackiw and S. Templeton, Phys. Rev. Lett. {\bf 48} (1982)
975;
Ann. of Phys. (New York) {\bf 140} (1982) 372.
\item{}C. Aragone, P.J. Arias and A. Khoudeir, Proceedings Silarg 7th, edited
by M. Rosenbaum et al, World  Scientific (1991) p.437.
\item{}C. Aragone and A. Khoudeir, Phys. Lett. {\bf 173B} (1986) 141; {\it
Quantum  Mechanics of Fundamental Systems}, ed. C. Teitlboin, (Plenum Press,
New York,  1988). p.17.
\item{}C. Aragone and S. Deser, Nucl. Phys. {\bf
B270}, [FS1] (1980) 329; \par M.A. Vasiliev, Sov. J. Nucl. Phys. {\bf 32}
(1980) 439. \item{}C. Aragone, S. Deser and Z. Yang, Ann. of Phys. {\bf 179}
(1987) 76. \item{}C. Fronsdal, Phys. Rev. {\bf D18} (1978) 3624.
\item{}C. Aragone and H. La Roche, Nuovo Cimento {\bf A72} (1982) 149.
\item{}T. Damour and S. Deser, Ann. Inst. Henri Poicar\'e {\bf 47} (1987) 277.
\item{}C. Aragone and A. Khoudeir, in preparation.
\item{}R. Jackiw and V. Nair, Phys. Rev. {\bf D43} (1991) 1942.
\item{}E.S. Fradkin and V. Y. Linetsky Ann. of Phys. {\bf 198} (1990) 293.
\item{}P.C. Argyres and C.R. Nappi, Phys. Lett. {\bf B224} (1989) 89.
\end{enumerate}
\end{document}